\documentclass[12pt]{article}

\usepackage{newtxtext,newtxmath}
\usepackage{booktabs}
\usepackage{braket}
\usepackage[utf8]{inputenc}

\usepackage{graphicx}

\usepackage[letterpaper,margin=1in]{geometry}

\renewenvironment{abstract}
	{\quotation}
	{\endquotation}

\date{}

\makeatletter
\renewcommand{\fnum@figure}{\textbf{Figure \thefigure}}
\renewcommand{\fnum@table}{\textbf{Table \thetable}}
\makeatother

\usepackage{scicite}

\usepackage{url}

\def\scititle{
	Attosecond charge migration timescales are dominated by transition dipoles, not correlations
}
\title{\bfseries \boldmath \scititle}

\author{
	Km Akanksha Dubey$^{\ast}$,
	Ofer Neufeld$^{\dagger}$ \and
    \small Technion - Israel Institute of Technology, Schulich Faculty of Chemistry, Haifa, 32000036, Israel.\and
	\small$^\ast$Corresponding author. Email: akankshadubey256@gmail.com\and
    \small$^\dagger$Corresponding author. Email: ofern@technion.ac.il
}

\begin{document} 

\maketitle

\begin{abstract} \bfseries \boldmath
Attosecond charge migration (CM) is an ultrafast process occurring when a molecule is irradiated by ultrashort laser pulses, creating a localized hole. The hole propagates rapidly through the molecule, generating electric currents and transferring charge across the molecular backbone. CM is a key ingredient in solar energy conversion, photosynthesis, and radiation damage. Despite its importance and intensive research, the fundamental physical and chemical mechanisms of CM remain not fully understood. Especially, a deeper insight into the role correlations play in the dynamics, and which chemical attributes determine CM timescales, is needed. Here we study with \textit{ab-initio} time-dependent density functional theory CM in the benchmark molecule, BrC$_4$H. We thoroughly explore CM under different initial conditions at the electronic and structural levels, including with theories of varying degrees of electronic correlations. We uncover a universal behavior where the hole moment (connecting to experimental observables) dominant frequency is roughly independent of all of these characteristics. In contrast, the timescales of the hole density evolution do vary with the chemical conditions and level of correlations. Employing a semi-analytical theory that reconstructs the hole moments in the cationic reference frame, we show that the attosecond timescale of CM is determined by molecular dipoles that filter out specific frequency responses with an analogy to optical selection rules. Our results provide essential insight into CM physics, which should be useful for interpreting attosecond experiments and engineering CM timescales by tailoring transition dipoles. 
\end{abstract}

\noindent
\section*{Introduction}
 Charge migration (CM) is a ubiquitous phenomenon that occurs when a molecule absorbs light, ionizing an electron, and leaving behind a positively charged cation. The hole in the cation then undergoes an ultrafast (attosecond-to-femtosecond timescale) motion along the molecular backbone, generating electric currents, and eventually de-cohering as energy dissipates into the electronic and vibronic systems  \cite{Cederbaum_CM1999,calegari_CM_2014,worner_2015science,martin_cm_cpl2017,Taran_NatPhys2026}. CM arises both naturally, e.g. as an initial stage in radiation damage due to X-ray light exposure in biosystems \cite{zewail_pnas2000,schuster_CT_DNA2000,amitava_radiation_damage_springer2007}, and routinely in laser-matter interactions where it can be experimentally probed and manipulated \cite{calegari_CM_2014,Matselyukh2022,worner_2015science,shaul_jacs_2023,Mansson_cmEXP_2021,Wanie_cmEXP_2024,calegari_EXP2018,He_cmEXP_2023,martin_cmEXP2021,lepine_cmEXP2019,trabattoni_cmEXP_rsta2019,weingartz_cmEXPtas_jpca2021,He_cmEXP_NatCommu_2022,Taran_NatPhys2026}. It also forms a decisive first step in charge transfer processes that are essential for applications such as photo-energy harvesting \cite{robb_CT_jcp2013,rozzi_ct_light-harvest2013_naturecommu,ct_science2014,Matselyukh_ct_exp2025}. Due to its great importance and abundance, CM has been intensively studied over the the last few decades, both experientially \cite{calegari_CM_2014,Matselyukh2022,worner_2015science,shaul_jacs_2023,Mansson_cmEXP_2021,Wanie_cmEXP_2024,calegari_EXP2018,He_cmEXP_2023,martin_cmEXP2021,lepine_cmEXP2019,trabattoni_cmEXP_rsta2019,He_cmEXP_NatCommu_2022,Taran_NatPhys2026} and theoretically \cite{nikolay_pnas2025,Folorunso2023,cederbaum_2013,Cederbaum_CM1999,photochem_2021,Yu_pra2023,gaarde_brc4h_jpca2024,tremblay_CM_endofull_jpcc2024,despre_2025ChemSci,shaul_jacs2022,martin_cm_cpl2017,martin_rsc2016,gaarde_pra2025,Vacher_prl2017}. However, despite substantial advancement, many questions remain. 
 
 Among these challenges, the role of electron-electron interactions (such as correlated dynamics) remains not fully resolved. Earlier seminal work in the field concluded that correlations are key in driving hole motion \cite{Cederbaum_CM1999,cederbaum_2013,Mansson_cmEXP_2021}, which is also supported by recent quantum chemical analysis \cite{despre_2025ChemSci,kuleff2025_pccp,kuleff_pra2022,kuleff_jpca2024}. 
However, recent works also qualitatively capture CM phenomena by applying time-dependent density functional theory (TDDFT) \cite{gaarde_brc4h_jpca2024,Folorunso2023,Folorunso_prl2021,despre_2025ChemSci,martin_cm_cpl2017,wu_cm_lda_pra2023} at semi-local exchange correlation (XC) density functional approximations \cite{gaarde_brc4h_jpca2024,despre_2025ChemSci,wu_cm_lda_pra2023,martin_cm_cpl2017,martin_rsc2016} that are a well known to capture only weak correlations. Implementations with hybrid functionals with portions of exact exchange are rare \cite{Folorunso2023,Folorunso_prl2021}, and a DFT benchmark study has not been performed as is commonly done for ground-state quantum chemistry problems \cite{garcia_dft_xc_benchmark_pccp2026} due to both numerical cost, and absence of quantitatively accurate experimental data. Generally, during the first femtosecond following photoionization when CM is initiated, many-body interactions are expected to re-normalize the electronic energy landscape, but it remains unclear how this process connects with the chemical attributes of the molecule or how crucial it is in the dynamics. 
 
The issue of correlations is intimately connected to a more tangible observable --- the CM timescale, which can be probed directly in attosecond pump-probe experiments \cite{calegari_CM_2014,Matselyukh2022,worner_2015science,Mansson_cmEXP_2021,Taran_NatPhys2026,He_cmEXP_2023,calegari_EXP2018,david_atas2022_sciadv,Barillot_PRX2021} and connects to the hole velocity. Since the hole motion is usually probed optically (e.g. with transient absorption \cite{Matselyukh2022,Taran_NatPhys2026,david_atas2022_sciadv}), measured data essentially captures the hole-induced dipole moments (polarization) over time. Past experiments observed a broad range of timescales across different types of molecular systems \cite{calegari_CM_2014,Matselyukh2022,worner_2015science,Mansson_cmEXP_2021,He_cmEXP_2023,Taran_NatPhys2026,calegari_EXP2018,david_atas2022_sciadv,Barillot_PRX2021}, roughly ranging from few hundreds of attoseconds up to few femtoseconds. It has been predicted that the velocity weakly depends on molecular size, but simultaneously depends on the bonding nature \cite{Folorunso_prl2021}. Additional systematic connections are generally missing and reflect our lacking knowledge of what/how chemical attributes determine charge migration characteristics. Resolving these challenges could pave way to exact chemical and optical manipulation of CM for emerging technologies. 

 Here we employ \textit{ab-initio} TDDFT to study CM in the benchmark molecule BrC\textsubscript{4}H. We perform an exhaustive exploration of this system with various levels of theory (comparing semi-local, hybrid, and meta XC functionals) and its chemical landscape (manipulating bond lengths, angles, and initial hole states). This thorough investigation leads us to several key findings. First, we show that correlations alone do not drive the process, nor are they essential in determining most physical attributes of CM such as its velocity or efficiency. Second, we show that the CM velocity and the timescale for charge transfer associated with it are very weakly dependent on most typical chemical features. Interestingly, despite the hole moments being independent of these features and correlations, the full hole density evolution considered across the entire molecular space does strongly depend on them. This creates a contradiction that we resolve by explicitly showing that transition dipoles in the cationic reference frame filter out most frequency components in the density evolution upon integration, leaving out only a single dominant frequency in CM hole moments. This situation resembles dipolar selection rules in optical excitations \cite{LandauLifshitzQM1977,Ofer_Rev_optica2026} and suggests that CM velocity is difficult to control by purely tuning the light source unless very drastic changes in frequency are employed. Instead, changes in bonding nature and strong alterations of the molecular orbitals are needed. Our work therefore opens a new path towards CM chemical manipulation.

\section*{Results}
\subsection*{\textit{Ab-initio} theory}

\begin{figure}
    \centering
    \includegraphics[width=0.9\linewidth]{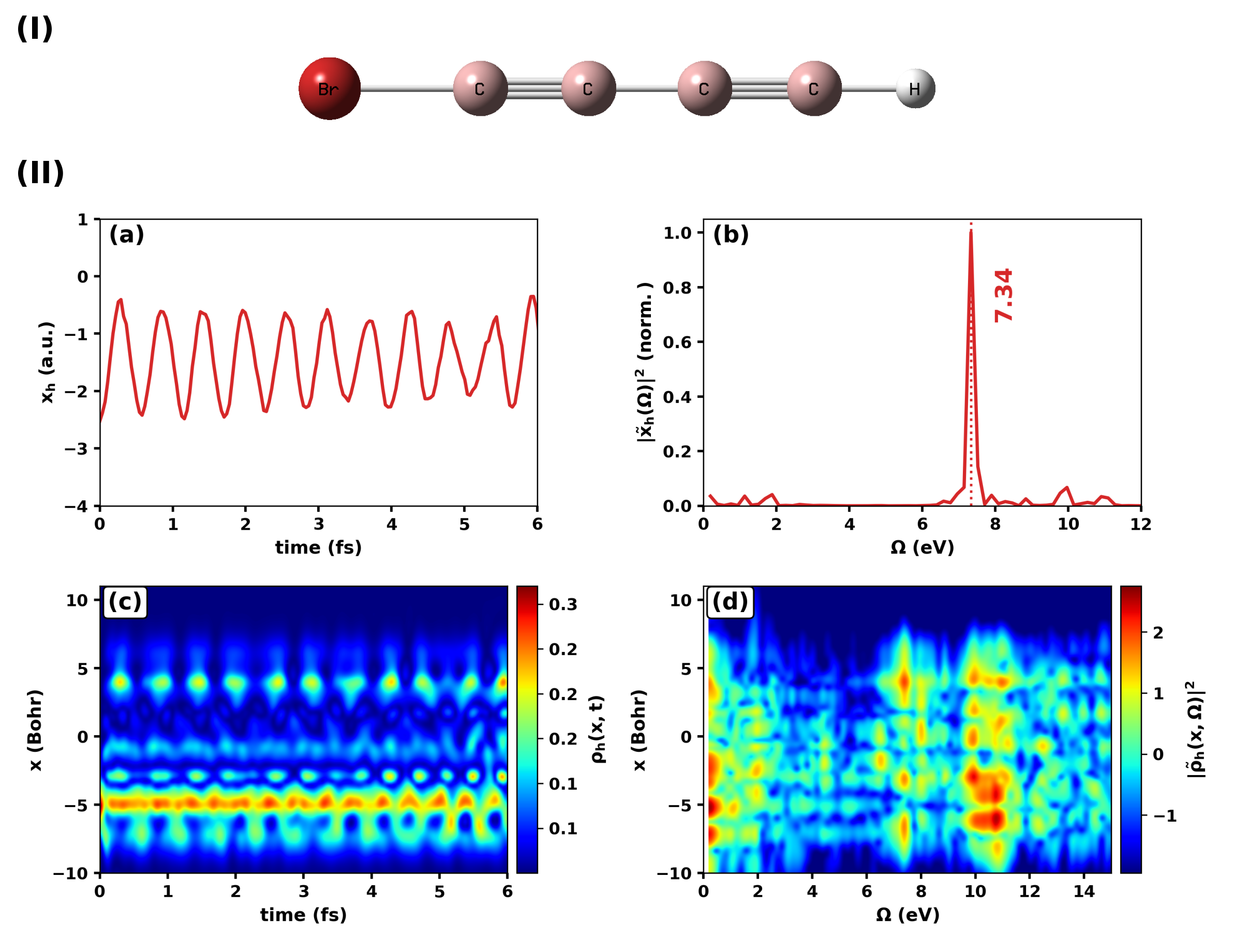}
    \caption{Panel (I): Geometry of BrC\textsubscript{4}H molecule. Panel (II): Dynamical CM evolution within the adiabatic LDA. (a) Temporal evolution the CM hole moment. (b) Spectral decomposition of hole moments from panel (a). (c) Spatio-temporal evolution of the hole density along the molecular axis. (d) Corresponding Fourier transform from (c).} 
    \label{fig:1}
\end{figure}

Let us begin by briefly describing our theoretical methodology. CM is described with \textit{ab-initio} TDDFT \cite{hardy_book_tddft2012,Ullrich_tddft_apl2025} on a real-space grid employing Octopus code \cite{octopus_jcp2020}. The molecule is aligned along the x-axis (see illustration in Fig.\ref{fig:1}(I)). To initiate CM we employ the sudden approximation (SA) \cite{sudden_approx_Barry_chemphys1977} assuming that the electron is pulled out instantaneously from the parent molecule, projecting the wave functions of the neutral onto the cation system. The electronic density is propagated numerically, from which we can extract the time-dependent hole density evolution, $\rho_h(x,t)$ (after integrating out the transverse inactive axes), as well as the hole moment $x_h(t)$ (describing the hole's average position along the molecular backbone). All additional details of the numerical scheme are delegated to the Materials and Methods section \cite{methods}. 

Starting out from a sudden ionization of the deepest valence state (denoted molecular orbital \#1 (MO1)) that is highly localized on the Br site, and employing the adiabatic local density approximation (aLDA) for the XC functional, we observe a rapid evolution of the hole density on attosecond timescales (with periodicity of $\sim$570 attoseconds, see Fig.\ref{fig:1} II(c)). The CM evolution is greatly similar to past results in this system, though the timescale slightly differs as our study covers CM in a higher energy regime where the hole is created from the innermost valence orbital rather than a deeper core state \cite{Folorunso_prl2021,gaarde_brc4h_jpca2024}. The four main observables key to our analysis are the hole density, $\rho_h(x,t)$, hole moment, $x_h(t)$, and their spectral decompositions in Fourier space, $\tilde{\rho}_h(x,\Omega)$ and $\tilde{x}_h(\Omega)$, respectively. Figure \ref{fig:1} as a whole presents one clear and outstanding result --- the hole moment evolves with a single dominant timescale associated with its periodic motion along the molecule, moving back and forth every $\sim$570 attoseconds (Fig.\ref{fig:1} II(a), associated with a frequency of $\sim7.34$ eV (Fig.\ref{fig:1} II(b)). On the other hand, the full density evolution involves a wide range of frequencies up to $\sim$15 eV. For instance, the most dominant frequency in the full density evolution is $\sim$11 eV (see Fig.\ref{fig:1} II(d), which is in logarithmic scale). This introduces a contradiction, and raises the question of where have the missing frequencies gone in the hole moment evolution, and why they have vanished upon integration.

\begin{figure}
    \centering
    \includegraphics[width=0.9\linewidth]{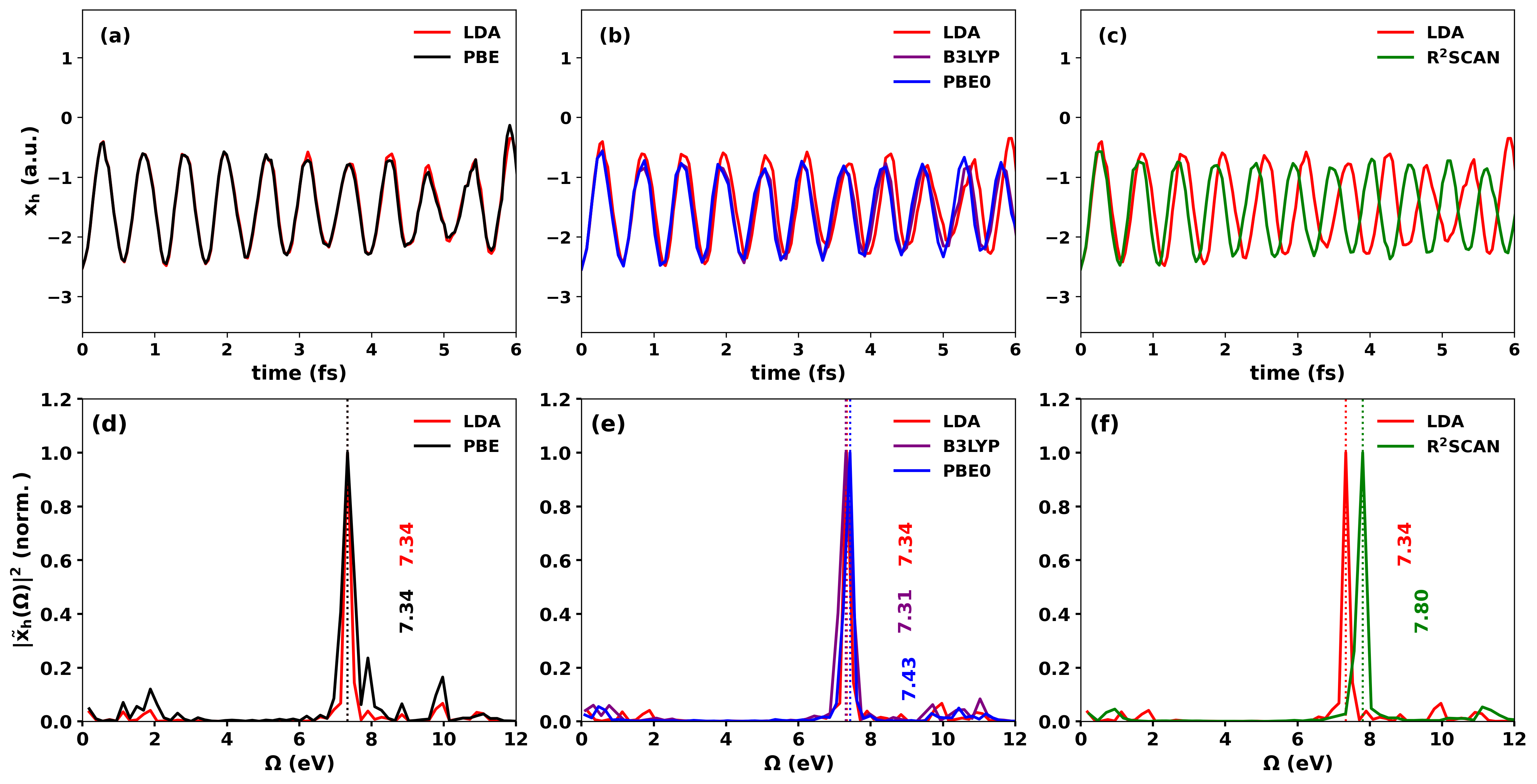}
    \caption{Quantitative comparison of hole moments and their respective spectral decompositions upon inclusion of different levels XC interactions. (a) Semi-local XC. (b) Hybrid XC. (C) Meta XC.} 
    \label{fig:2}
\end{figure}

To further explore this question, we repeat the above simulations by employing different types of XC approximations, including semi-local, meta, and hybrid, functionals. Our motivation is that since CM has been in the past identified as correlation driven, comparison of these different levels of theory might uncover if correlated dynamics determine the missing frequencies in the hole moments, and the CM emerging timescale. Figure \ref{fig:2} shows that this hypothesis is incorrect. Remarkably, the dominant hole moment frequency and timescale is essentially independent of the correlation-level in the theory. Indeed, the largest shift occurs with R$^2$SCAN \cite{r2scan_furness_jpcl2020}, and is only a minute 6\% change in timescale. This unambiguously shows that correlations do not dominate CM, nor do they determine its most basic property --- its velocity.  

\begin{figure}
    \centering
    \includegraphics[width=0.9\linewidth]{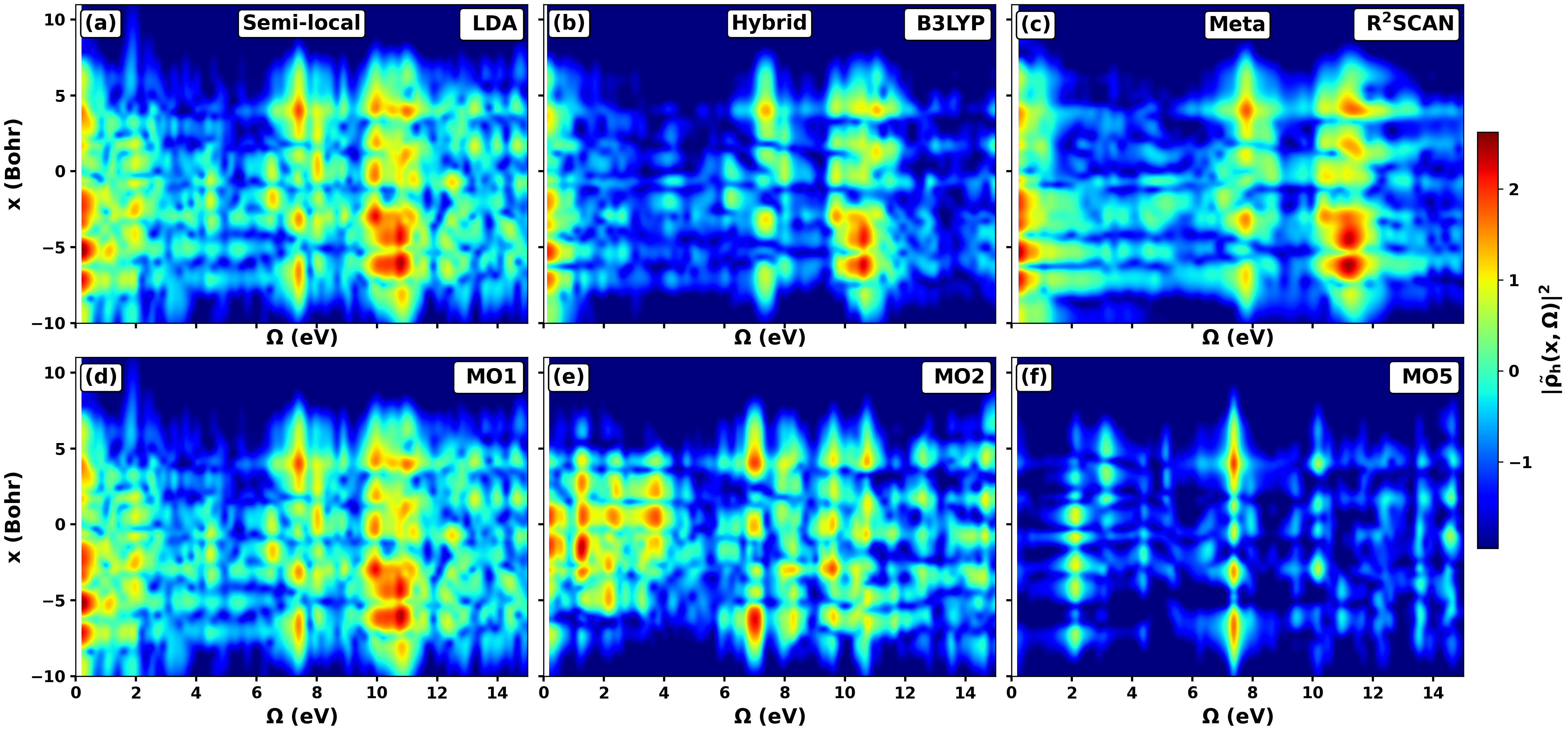}
    \caption{Hole density evolution elucidating the spectral content of the full hole denisty (not hole moments), for select XC functionals (a-c). (d-f) Same as in (a-c), but for different choices of initial ionized molecular orbital with varying degrees of localization (see text).}
    \label{fig:3}
\end{figure}

Contrarily to this result, Figure \ref{fig:3}(a-c) shows the full spectral content of the hole evolution, $\tilde{\rho}_h(x,\Omega)$ for select XC cases (other XC functional cases are delegated to the Supplementary Materials (SM)). Here the level of theory plays a major role and substantially alters the main spectral lines. This is the more expected physical result considering that upon changing the XC functional level the orbital eigen-energies themselves change, and there is no reason to expect similar CM attributes. Thus, the apparent mismatch between the hole moment and hole density evolutions appears as a universal result, whereby hole moments are insensitive to correlations, unlike the full hole density.

\begin{figure}
    \centering
    \includegraphics[width=0.9\linewidth]{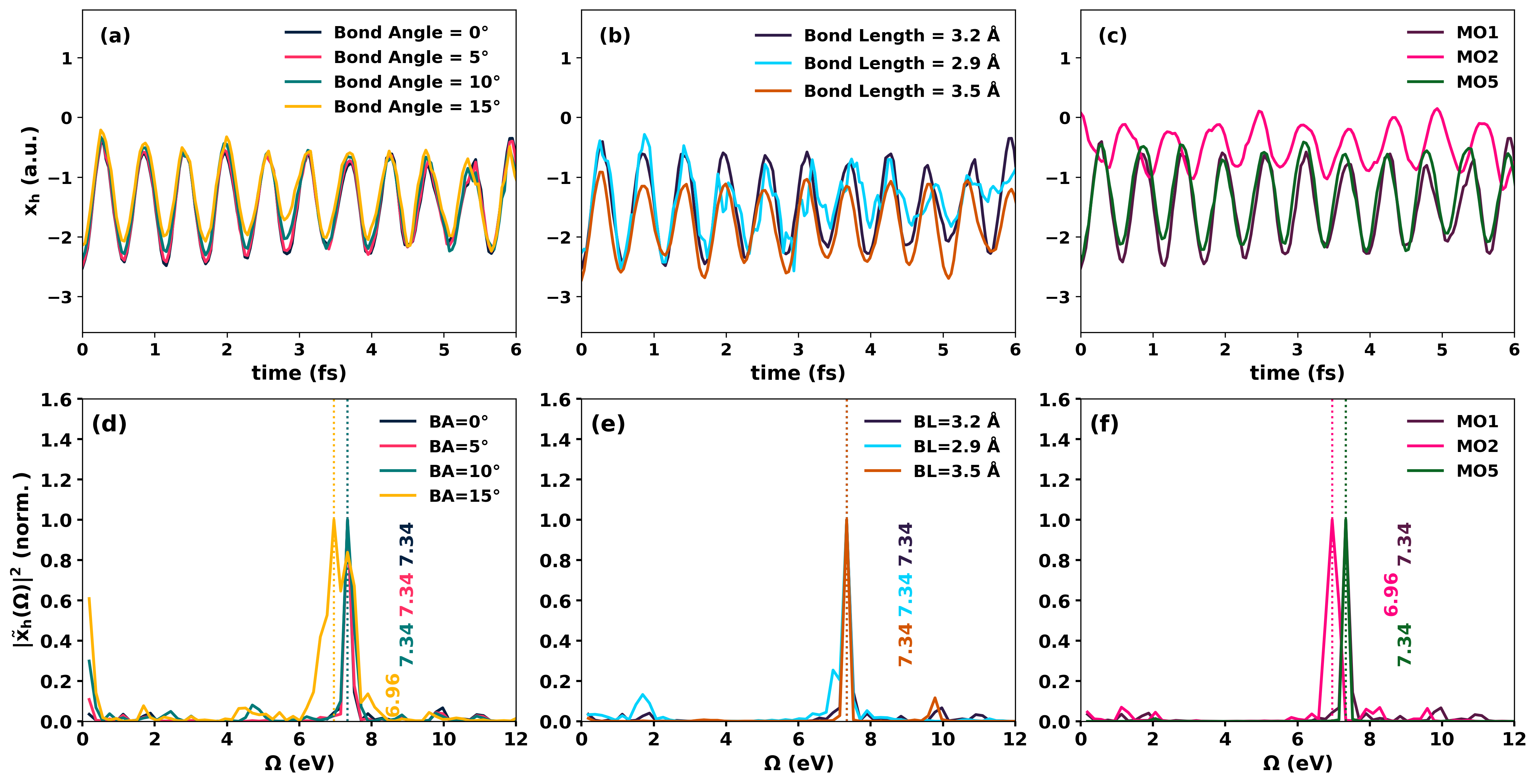}
    \caption{CM hole moment analysis by variation of chemical attributes of the molecule. (a-c) Temporal evolution of the hole moment on attosecond to femtosecond scale considering various molecular attributes such as Bond Angle (BA), Bond Length (BL) and Molecular Orbital (MO), variations. (d-f) Corresponding spectral content of hole moments for each varied molecular attribute.}
    \label{fig:4}
\end{figure}

This motivates us to explore more fully the chemical landscape within this benchmark system. Figure \ref{fig:4} shows hole moments and the emerging spectrum for three types of chemical modifications: (i) The bond angle (BA) between Br and its neighboring C atom out of the molecular plane in increments of $5^\circ$, which for the optimized original geometry lies at $0^\circ$. (ii) Bond Length (BL) variations between the Br atom and neighboring C atom in increments of 0.3 $\mathrm{\AA}$, where the original position for the Br atom is $\sim$3.2 $\mathrm{\AA}$. (iii) Considering orbital-selected electron ionization from different MOs that are not as strongly localized on the Br site. Figure \ref{fig:4}(a-c) presents hole moments as they evolve once an electron is ionized under these various molecular attributes. The variation in bond angle (Fig. \ref{fig:4}(a,d)) hardly modifies the hole moment evolution and the oscillatory structure remains intact over the temporal scale of $\sim$6 fs. Contrarily, the modification in bond-length induces slight variations in the hole moment evolution --- mostly in its slower response and dissipation (Fig. \ref{fig:4}(b)). Nonetheless, the dominant frequency and CM velocity is virtually unchanged (Fig. \ref{fig:4}(e)), despite the hole needing to transverse longer/shorter distance across the Br-C bond. This suggests that the molecular system internally compensates somehow for the change in distance in this local region to fix the hole velocity across the whole molecule. Finally, we analyze in Fig. \ref{fig:4}(c) CM initiated by MO1, MO2, and MO5, which have completely different localization properties. Indeed, only MO1 is strongly localized on the Br site, while MO2 has weaker localization, and MO5 is nearly de-localized across the entire molecule (see SM). Figure \ref{fig:4}(c) shows almost perfect overlap of hole moments for MO1 and MO5 up to $\sim$6 fs, whereas hole moment in MO2 evolves with a different magnitude as well as a slight shift in phase throughout. Overall though, the hole moment timescale associated with the hole velocity is basically untouched (Fig. \ref{fig:4}(f)). This result is striking and highly unexpected. Indeed, different ionized orbitals should lead to completely different CM characteristics as the ionized orbital pre-defines the superposition of cationic states that inevitably lead to CM. In general, the results thus far propose a curious puzzle as to why these different molecular attributes and level of correlations lead to almost a similar temporal profile for the hole moment.

Next, we analyze the corresponding spectral content of the full hole evolution in the molecule for all of these cases. Most cases are delegated to the SI, while the most striking variation in MO is displayed in Fig.\ref{fig:3}(d-f). The contrast from Fig.\ref{fig:4}(c,f) couldn't be larger --- changing the initial ionized orbital strongly alters $\tilde{\rho}_h(x,\Omega)$. For instance, in the case of MO2 (Fig.\ref{fig:3}(e)) dominant low-energy frequencies emerge in the range of 1-4 eV that are non-existent in the MO1 case. Higher energy modes in 8-12 eV range are similarly much more apparent. For MO5 (Fig.\ref{fig:3}(f)), nearly the entire spectrum comprises the dominant $\sim$7.34 eV peak, and other peaks are negligible in comparison (unlike in MO1 where many other frequencies play a role and in fact the most dominant frequency is at $\sim$10.5 eV). The situation for the bond angle and length variation is quite similar, where variation in the chemical attributes induces variations in the hole evolution spectra (see SM). Inevitably though, these alterations do not translate into strong modifications in the CM hole moments. This result essentially generalizes the behavior seen above in the case of LDA alone (Fig.\ref{fig:1}) and various XC functionals (Fig.\ref{fig:2} - Fig.\ref{fig:3}). 

This is a crucial element in our analysis, since the hole moments connect with the molecular dipoles and in fact are the entities that could be measured in pump-probe experiments that lack spatial resolution (e.g. in attosecond transient absorption spectroscopy \cite{Matselyukh2022,Taran_NatPhys2026,david_atas2022_sciadv}). We note that in our analysis, we eventually do see alteration also in hole moment timescales if the variation in molecular attributes strong enough. For instance, by initiating CM from the HOMO and HOMO-1 orbitals that are grossly delocalized across the molecule, the hole moment dominant timescales shift to $~\sim$1.3 fs (see SM). Nonetheless, drastic changes are required to cause this modification. At face value, this result helps justify why clear CM signatures have been observed even in the strong-field ionization case where multiple high-lying orbitals are involved \cite{Matselyukh2022,He_cmEXP_2023}, as well as with attosecond pulses with  broad bandwidth that ionize multiple states \cite{Taran_NatPhys2026,Mansson_cmEXP_2021,david_atas2022_sciadv}. In the following section, we attempt to construct a semi-analytic theory explaining this result by analyzing CM in the cationic reference frame.

\subsection*{Cationic dipolar CM reconstruction}

\begin{figure}
    \centering
    \includegraphics[width=0.8\linewidth]{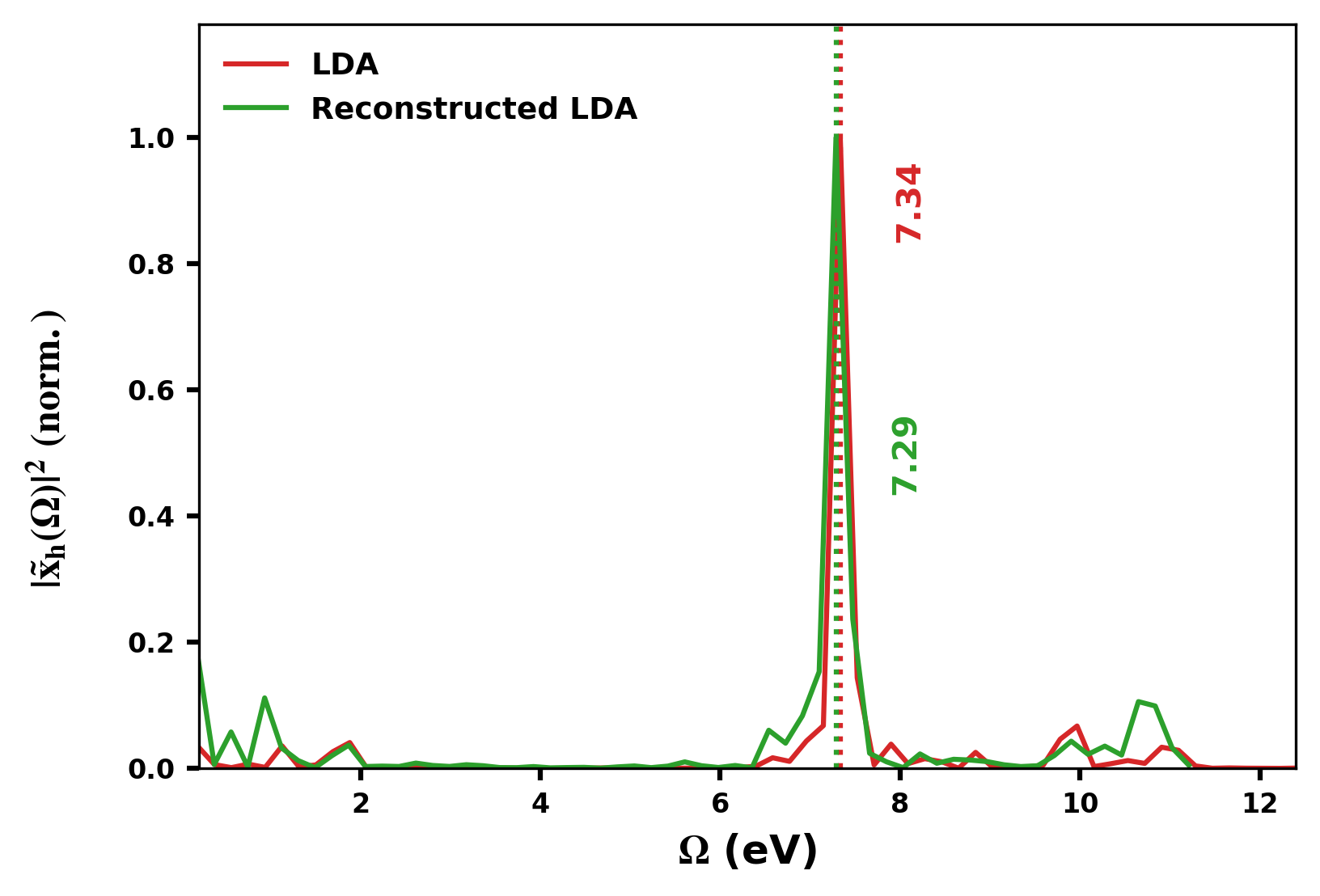}
    \caption{Spectral decomposition of hole moments in CM within adiabatic LDA, employing \textit{ab-initio} simulation directly (as in Fig. 1), from the semi-analytical reconstruction approach that assumes negligible electronic correlations (see text).}
    \label{fig:5}
\end{figure}

Employing a semi-analytical approach, we reconstruct the hole moment from the evaluated density matrix (coherences and occupations) and transition dipole matrix elements, all of which are extracted from the \textit{ab-initio} simulations (either at the ground state, or extracted from the dynamical evolution). We start by spanning the electronic orbital in the neutral basis set as a coherent superposition of cationic system orbitals that form a complete orthogonal basis, which can be written as:
\begin{equation}
\left|\psi^+_J(t)\right\rangle
=
\sum_{I} C_{IJ}(t)\,
\left|\phi^{+}_{I}\right\rangle. \label{1}
\end{equation}
In eq. \ref{1} $\left|\psi^+_J(t)\right\rangle$ denotes the evolving orbitals in the cationic system, but written in the neutral system orbital basis set that propagates and solves eq. \ref{3} (time-dependent propagation of the CM dynamics). $\{\left|\phi^{+}_{I}\right\rangle\}$ is the complete set of cationic orbitals upon which the KS states of the neutral system are projected onto, which can be separately obtained by a ground state calculation (the KS superscript is hensceforth dropped). In equation \ref{1}, $C_{IJ}(t)=c_{IJ}(t)e^{-iE^+_{I}t/\hbar}$, is the coefficient of occupation of individual orbitals in the cationic system, including the full temporal evolution (most of which arises from the eigen-energy, $E^+_{I}$, freely-evolving system as there is no external perturbation). Any deviations from the free evolution could be taken to arise as a result of many-body interactions that could in principle re-normalize the orbitals themselves (causing $c_{IJ}(t)$ to be a time-dependent function) and eigen energies (changing $E^+_{I}$ into $E^+_{I}(t)$) in the cationic system. We obtain the occupation coefficient of the individual cationic orbitals as $C_I(t)=\sum_{J} C_{IJ}(t)$. The full coefficients are computed by projecting the time-dependent evolved KS orbitals onto the cationic GS KS basis, $C_{IJ}(t)=\langle\phi^+_I\mid\psi^{+}_J(t)\rangle$, arising from the ab-initio dynamical simulation. 
We further compute coherences and occupations in the cationic reference frame by defining $\eta_{IJ}(t)=C_I^{*}(t)C_J(t)$, and $\eta_{II}(t)=|C_I(t)|^2$, for reasons that will become apparent below. Subsequently, the hole moment $x_{h}(t)$ can be reconstructed from the coherences and occupations via the relation:
\begin{equation}
\begin{split}
d_{h}(t)
= \langle \sum_I \phi^+_I(t)\mid \hat{x} \mid \sum_J \phi^{+}_J(t)\rangle \\
= \langle \sum_I C^*_I(t)\phi^+_I\mid \hat{x} \mid \sum_J C_J(t)\phi^{+}_J\rangle \\
=\sum_{J,I} \eta_{JI}(t)\mu_{IJ}. \label{2}
\end{split}
\end{equation}
Here $|\phi^+_I(t)\rangle$ are in principle the actual evolving KS orbitals of the cationic system with the hole dynamics, which are allowed to have time dependence in the case that many-body correlations alter the orbitals and eigenenergies during evolution (as discussed above). In that case they can be represented as a sum of states of the ground state orbitals that are a complete basis as well. In practice, we perform an approximation that assumes the orbitals themselves change negligibly over time, hoping that correlations do not play a major role. This reduces one sum and greatly simplifies the analysis. $\mu_{IJ}$ in equation \ref{2} denotes the molecular transition dipole moments between coherent cationic ground state orbitals $I$ and $J$, which are separately calculated as $\mu_{IJ}=\langle \phi^+_I\mid \hat{x}\mid \phi^{+}_J\rangle$. This further assumes that $\mu_{IJ}$ has negligible time-dependence such that correlations weakly alter the orbitals, in similar spirit to the approximation employed above. Note that $d_{h}(t)$ is not entirely identical to the \textit{ab-initio} calculated $x_{h}(t)$ due to the various approximations employed, but, if correlations are weak they are expected to be very close. 

Employing this approach to the case of the aLDA XC functional from Fig.\ref{fig:1}, we see that the CM hole moment dominant frequency is fully reconstructed (Fig.\ref{fig:5}). This is expected in the sense that the full chemical evolution is embedded in the dipolar and occupation/coherence terms that are extracted \textit{ab-initio}. Still, the quality of reconstruction is surprising in the sense that eqs.\ref{1}-\ref{2} make a strong assumption regarding dynamical correlations --- they are neglected by not allowing the cationic states to re-normalize during the evolution. In principle, the dipolar transition elements should vary over time since the cationic orbitals themselves slightly change. Only partial impact of interactions are considered in Fig.\ref{fig:5}, by allowing dynamical tuning of the energy of the orbitals and their occupations in the projection coefficients that are generally complex. 

Further analyzing the semi-analytical calculations, we find that pairs of $I$ and $J$ cationic orbitals (see equations \ref{1} to \ref{2} including mathematical expressions in between) contribute to significant density matrix evolution at varying frequencies (i.e. to $\tilde{\rho}_h(x,\Omega)$). The contributions are focused around the main transition frequencies in the cation $\Delta_{IJ} = E_I-E_J$, suffice the states are occupied by the hole. This gives rise to the broad spectral content in $\tilde{\rho}_h(x,\Omega)$ that is also sensitive to correlations and molecular geometry (since those slightly alter the frequencies, and largely alter occupations of states). However, many of these sets of pairs $(I,J)$ are filtered out by the corresponding cationic dipole matrix elements $\mu_{IJ}$ (see eq. \ref{2}, leading to only dipole-selected specific pairs that survive in the full eq. \ref{2}, and additional discussion in the SM). Thus, these specific pairs, though selected differently in each XC functional and molecular attributes cases (evident from the full hole density spectral content in Figs. \ref{fig:1} and \ref{fig:3}), eventually form a similar dominant frequency. Thus, occupations of the cationic states plays only in a minor role as long as not too drastic differences in initial conditions are taken (e.g. by occupying initial states in completely separate energy landscapes, see SM). The marks the main physical player in the dynamics of hole moments as the molecular cationic dipoles that govern the observed universal behavior of CM hole moments, at least in this benchmark linear molecule. This situation resembles filtration due to optical selection rules, e.g. as in a quantum harmonic oscillator. One can then imagine the molecule as a big linear well, and the hole moves back and forth within the well. Even if one occupies a forest of initial superposition states in the harmonic oscillator, only a single frequency contributes to the dipole moment ($\tilde{x}_h(\Omega)$) due to optical selection rules that forbid transitions by more than one energy level ($\Delta n=\pm1$). This is despite the fact that the full density evolution ($\tilde{\rho}_h(x,\Omega)$) in the well will indeed exhibit all of the eigen-energies associated with the occupied states, $\Delta_{IJ}$. The situation in the molecule is very similar, except that it's not strict selection rules that limit the main frequency, but just the typical available dipole transition with strong oscillatory strength (see SM).

This analysis uncovers that the dipolar elements are essential in CM. Notably, they arise directly from the molecular geometry and bonding nature itself, i.e. from the orbitals of the cation. Overall, our results show that cationic dipoles play the main role in CM timescales and hole moment evolution, while correlations and molecular finer details tune the complete hole evolution in a spatially-resolved manner that is not always accessible in experiments. Ultimately, correlations and many-body interactions negligibly impact hole dynamics and the CM timescales.

\section*{Discussion}
To summarize, we analyzed attosecond charge migration induced by a sudden ionization event in the benchmark molecule BrC\textsubscript{4}H using \textit{ab-initio} adiabatic TDDFT. By employing simulations with different levels of electronic interactions and analyzing dependence on the molecular geometry, we were able to uncover general trends that govern the CM timescales. First, we showed that correlation are in no way necessary to initiate the process, and in practice, carry little impact on the hole moment dominant frequency (its velocity). This result formally justifies application of weakly-correlated semi-local XC functionals for capturing CM, at least in similar nature organic systems. Similarly, minor changes in molecular geometry (without changing the bonding nature) only marginally tune the CM velocity. Remarkably, even changes in the initial hole state do not strongly impact the dominant CM timescales as long as those changes are in a not too broad energy window. This result helps justify CM observations even under strong fields \cite{Matselyukh2022,He_cmEXP_2023} and broad pump bandwidths \cite{Mansson_cmEXP_2021,david_atas2022_sciadv}. In contrast, the full spectral evolution of the hole density does depend on these molecular attributes, but this sensitivity ends up not translating to changes in the hole moment temporal evolution. By expanding a semi-analytic theory in the cationic reference frame, we showed that this contradiction arises due to dipolar transition elements that filter out most frequency components of the full density evolution when calculating the hole average position. This result is in close analogy to optical selection rules in a harmonic oscillator. 

Taken together, our results pinpoint molecular dipoles as the prominent source that control timescales in attosecond CM, not electronic correlations. While our study was specific to a benchmark linear organic molecule, we expect the main conclusions to be largely transferable due to their fundamental nature. Looking ahead, our analysis should motivate novel experiments for controlling CM by chemically and optically tuning molecular dipoles, as well as to perform spatially resolved experiments that could track down the missing CM frequencies.


\clearpage 

%
\bibliography{manus} 
\bibliographystyle{sciencemag}

%
%
%
%
%
%


\section*{Acknowledgments}
O.N. gratefully acknowledges the scientific support of Prof. Dr. Angel Rubio and the Young Faculty Award from the National Quantum Science and Technology program of Israel’s Council of Higher Education Planning and Budgeting Committee. O.N. and K.A.D. gratefully thank the Technion Helen Diller Quantum Center for financial support.
\paragraph*{Author contributions:}
K.A.D and O.N. contributed equally to this work.
\paragraph*{Competing interests:}
There are no competing interests to declare.
\paragraph*{Data, code and materials availability:}
The data related to this work are presented in the article and Supplementary Materials. Octopus code is an open-source software that can be freely accessed from its official website.


\subsection*{Supplementary materials}
Materials and Methods\\
Supplementary Text\\
Figs. S1 to S4\\
Tables S1\\
References \textit{(48-\arabic{enumiv})}\\ 


\newpage


\renewcommand{\thefigure}{S\arabic{figure}}
\renewcommand{\thetable}{S\arabic{table}}
\renewcommand{\theequation}{S\arabic{equation}}
\renewcommand{\thepage}{S\arabic{page}}
\setcounter{figure}{0}
\setcounter{table}{0}
\setcounter{equation}{0}
\setcounter{page}{1} 


\begin{center}
\section*{Supplementary Materials for\\ \scititle}

Km Akanksha Dubey$^{\ast\dagger}$,
Ofer Neufeld$^\dagger$\\
\small$^\ast$Corresponding author. Email: akankshadubey256@gmail.com\\
\small$^\dagger$Corresponding author. Email: ofern@technion.ac.il\\
\small$^\dagger$These authors contributed equally to this work.
\end{center}

\subsubsection*{This PDF file includes:}
Materials and Methods\\
Supplementary Text\\
Figures S1 to S4\\
Tables S1\\


\newpage


\subsection*{Materials and Methods}
\subsection*{Ground state calculations}
We start with detailing methodology used in ground state (GS) \textit{ab-initio} calculations for our model system, BrC\textsubscript{4}H. For this purpose we employ density functional theory (DFT) within the local density approximation (LDA) exchange-correlation (XC) functional with an added self-interaction correction (SIC) \cite{SIC_2002} to accurately account for long-range interaction of the XC functional (we discuss LDA case in detail here, later on theoretical discussion will be updated in terms of different XC functionals only). We employ the real-space grid-based Octopus code implementation of the DFT and time-dependent DFT to perform our simulations \cite{octopus_jcp2020}. The KS equations were discretized on a Cartesian grid with a spherical simulation box of radius 32 Bohr (a.u.). The selected grid spacing is 0.3 Bohr along all the axes. The ground-state KS equations are solved self-consistently for each case with a self-consistent field (SCF) threshold of 10$^{-8}$ Hartree. We employed the average-density SIC \cite{SIC_2002} scheme to account for self-interaction correction  (a SIC was not added for hybrid and meta functionals). For the core states, we employ frozen core approximation using norm-conserving pseudopotentials \cite{norm-cons_pseudo_1998}. When treating different molecular attributes, we begin with a new GS calculation for each case along with their respective optimized geometry (wherever required). The GS calculation provides the neutral system GS density and starting point for the time-dependent (TD) coherent evolution of the system upon electron ionization. Atomic units (a.u.) are used throughout, unless specified otherwise. 

\subsection*{Hole creation}
We adopt the sudden ionization approximation where the ionization step is assumed to be instantaneous and abrupt \cite{sudden_approx_Barry_chemphys1977}. An electron is removed from an inner orbital in the neutral system, generating the initial hole state. The ionized system forms a coherent superposition of cationic states in the sense that the ionized neutral orbital comprises a superposition of orbitals in the cationic reference frame. In order to study the role of varied degrees of hole localization, we selectively choose different molecular orbitals (MOs)- MO1, MO2, and MO5, in their decreasing extent of localized hole on the Br site (see Fig. \ref{fig:si3}). 

\subsection*{Time-dependent propagation of CM}
Time evolution of the entire system is studied using \textit{ab-initio} TDDFT in the adiabatic approximation for the XC functional, governed by time-dependent Kohn-Sham (TDKS) equations as-
\begin{equation}
i \partial_t \ket{\psi^{+KS}_i(\mathbf{r},t)} =
- \frac{1}{2} \nabla^2 \ket{\psi^{+KS}_i(\mathbf{r},t)} + \left[V^+_{KS}(\mathbf{r},t)\right]\ket{\psi^{+KS}_i(\mathbf{r},t)}. \label{3}
\end{equation}
The KS potential $V^+_{KS}(\mathbf{r},t)$ can be further decomposed as: 
\begin{equation}
V^+_{KS}(\mathbf{r},t)=V^+_{ion}+\int d^3r' \frac{\rho^+(\mathbf{r'},t)}{|\mathbf{r}-\mathbf{r'}|}+V^+_{XC}[\rho^+(\mathbf{r},t)]. \label{4}
\end{equation}
Here, $\ket{\psi^{+KS}_i(t)}$ is the $i^{th}$ time-dependent KS orbital of the cation system (after electron removal), but still in the basis set of the neutral system. Further, $V^+_{ion}$ accounts for the electronic interaction with the nucleus and core electrons (in the cation system). The second term in eq. \ref{4} accounts for the Hartree potential where $\rho^+(\mathbf{r},t)$ is the time-dependent electron density after sudden ionization, and $V^+_{XC}[\rho(\mathbf{r},t)]$ denotes the corresponding XC potential in the adiabatic approximation. 
From the time-dependent electron density, we further compute hole moment as follows-
\begin{equation}
    x_{h}(t) = \int x \rho_h(x,t) \Theta[\rho_h(x,t)] \ dx. \label{5}
\end{equation}
Here $\rho_h(x,t)$ is the hole density along the molecular backbone that is directly evaluated from the difference of cationic time-dependent electronic density and ground-state density and integrating out the transverse inactive coordinates. That is, we obtain $\rho_h(\textbf{r},t)=\rho^+(\textbf{r},t)-\rho_0(\textbf{r})$, where $\rho^+(\textbf{r},t)$ is the electron density of the cationic system that comprises the sum of occupied KS orbitals being propagated following ionization ($\rho^+(\textbf{r},t)=\sum _j |\psi^{+KS}_j(\mathbf{r},t)|^2$), and $\rho_0(\textbf{r})$ is the electron density of the GS neutral system prior to ionization. From $\rho_h(\textbf{r},t)$ we obtain the hole density along the backbone as: $\rho_h(x,t)=\iint{\rho_h(\textbf{r},t)} dy dz$. In eq. \ref{5}, $\Theta$ is a step function that further selects only positively valued density regions (holes) and makes sure $x_{h}(t)$ actually tracks the hole position in space such that negatively charged density regions with accumulation of electrons do not cause major cancellations and impact the integral's evaluation of the hole position. 
In order to obtain the spectral intensities, we Fourier transform the $x_{h}(t)$ obtained from equation \ref{5}, and similarly for $\rho_h(x,t)$ along each position coordinates.
Ion motion is neglected within the fixed-nuclei approximation that expected to be valid on these timescales ($<$10 fs). The time step for all time-dependent propagation is taken as 0.075 a.u. ($~\sim1.8 \ as$) for all cases, except for hybrid and meta XC functionals where it is taken as 0.15 a.u. ($~\sim3.6 \ as$) where self-consistent propagation is employed within Octopus code (converging the propagator in each time step up to $10^{-5}$ relative normalized tolerance). All time-dependent simulations are tested for convergence.



\subsection*{Supplementary Text}
The description below contains additional \textit{ab-initio} simulation results and semi-analytical analysis not shown in the main text. 

\subsection*{Additional \textit{ab-initio} simulation results}
\subsubsection*{Spectral decomposition of evolving hole density for various theory levels and molecular attributes}




\begin{figure}
\centering
\includegraphics[width=0.9\textwidth]{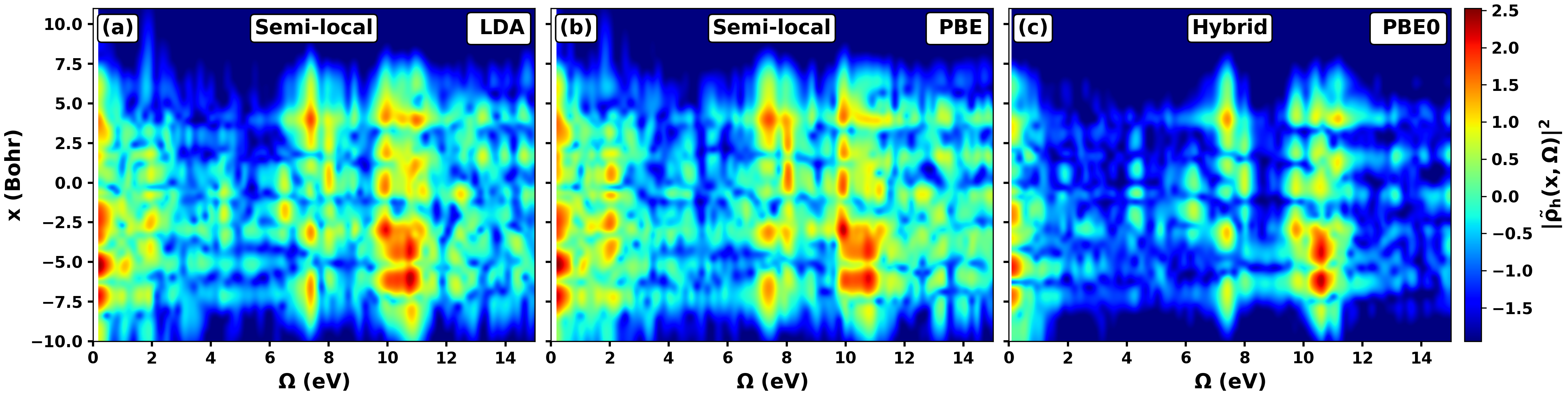}
\caption{Fourier-transformed hole density profiles showing spatial distribution of the spectral content simulated employing different theory levels for accounting electron-electron correlations- semi-local (a-b), and hybrid (c) XC functionals.}
 \label{fig:si1}
\end{figure}
Fig. \ref{fig:si1}(a-c) presents spectral content of the hole density evolving in our selected molecular system BrC\textsubscript{4}H upon sudden ionization from the inner valence molecular orbital MO1 considering different exchange-correlation (XC) functionals in our time-dependent density functional theory (TDDFT) CM simulations- (a-b) semi-local (LDA and PBE), and (c) hybrid (PBE0) functionals. It is evident from Fig.\ref{fig:si1}(a-b) that hole density evolves with similar spectral decompositions for both the semi-local XC functionals. The highest spectral intensity is seen centered within two regions around- 7-8 and 10-12 eV; followed by a uniform less intensity region beyond 12 eV. On the other hand, the hybrid functional in Fig.\ref{fig:si1}(c) presents a sharp contrast to semi-local hole density spectral contents, where spectral contents are missing at lower frequencies, but the previously two marked regions of highest intensity are preserved with a lesser intensity and width. Interestingly, the spectral content beyond 12 eV timescale is highly suppressed. Fig.\ref{fig:si1} complements spectral decomposition of hole density profiles (for the remaining cases) not shown in the Figure 3 in the main text. 

\begin{figure}
\centering
\includegraphics[width=0.9\textwidth]{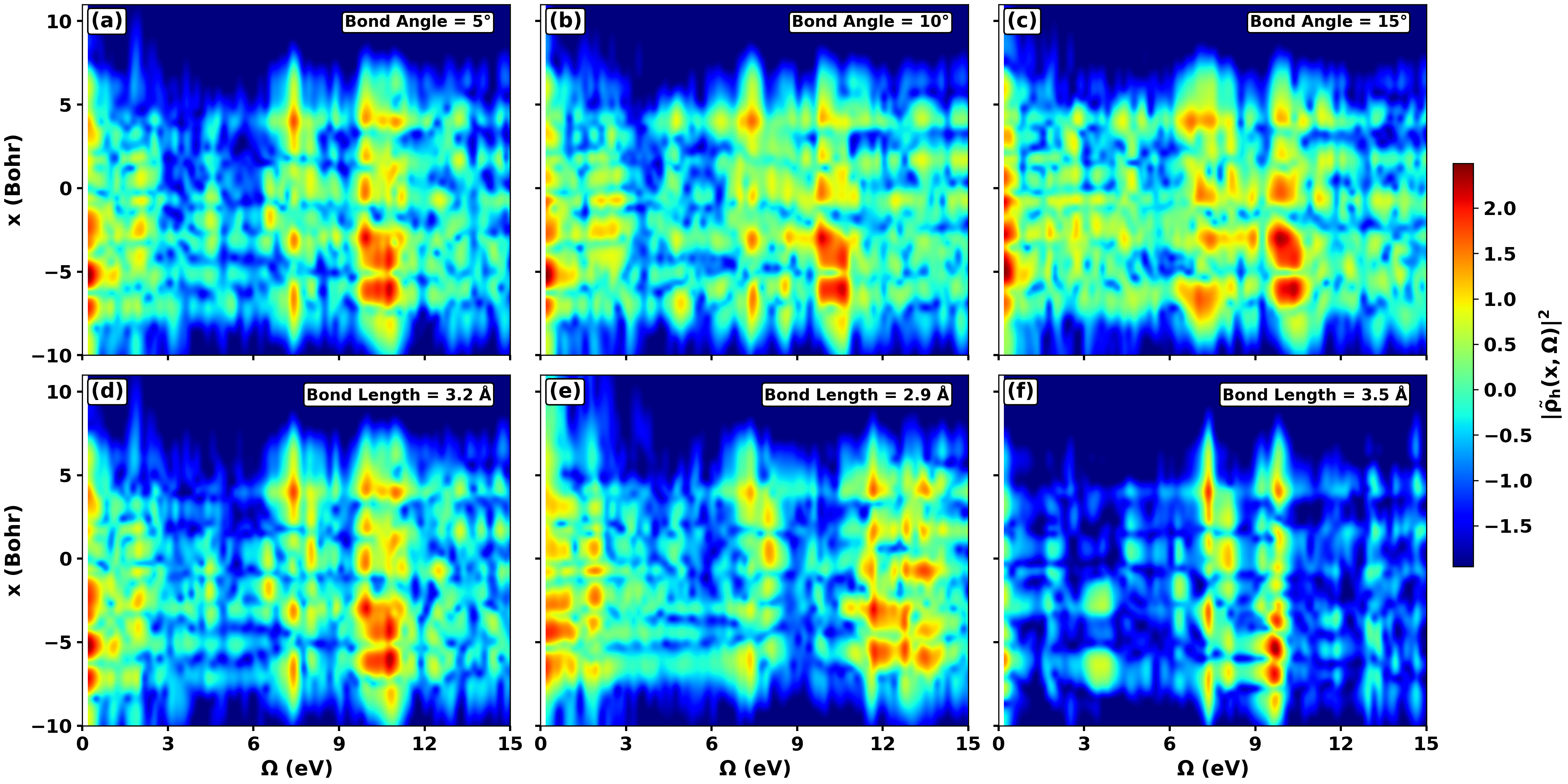}
\caption{Spectral decompositions of hole density profiles simulated employing various molecular attributes: (a-c) Bond angle (between Br and neighboring C atom), (d-f) Bond length (between Br and neighboring C atom) variations.}
\label{fig:si2}
\end{figure}

Evolution of the spectral contents of the hole density upon considering various molecular attributes are presented in Fig.\ref{fig:si2}. Fig.\ref{fig:si2}(d) shows the Fourier-transformed hole density of the molecule corresponding to its original optimized geometry- which is a Bond Angle=$0^\circ$ case. We notice that varying the bond angle from $0^\circ$ (d) to $5^\circ$ (a), the spectral distribution remains almost the same, with two marked regions of the highest intensity around 7-8 and 10-12 eV. Further increasing the bond angles in the transverse direction to the molecular backbone in (b-c), it is seen that new spectral contents arise increasingly between the region 3-6 eV and the maximum intensity region between 7-8 eV broadens. As we decrease the bond length (e) from the original optimized geometry (d), the highest intensity spectral contents located within 10-12 eV vanish and rather a similar region appears beyond 12 eV. Moreover, as we increase the bond length (f), interestingly the widely spread spectral contents are squeezed into the two regions of the highest intensity, which is further narrowed down- specifically the 10-12 eV region. Fig.\ref{fig:si2} complements the Figure 5 in the main text corresponding to figure 4. 

\begin{figure}
\centering
\includegraphics[width=0.9\textwidth]{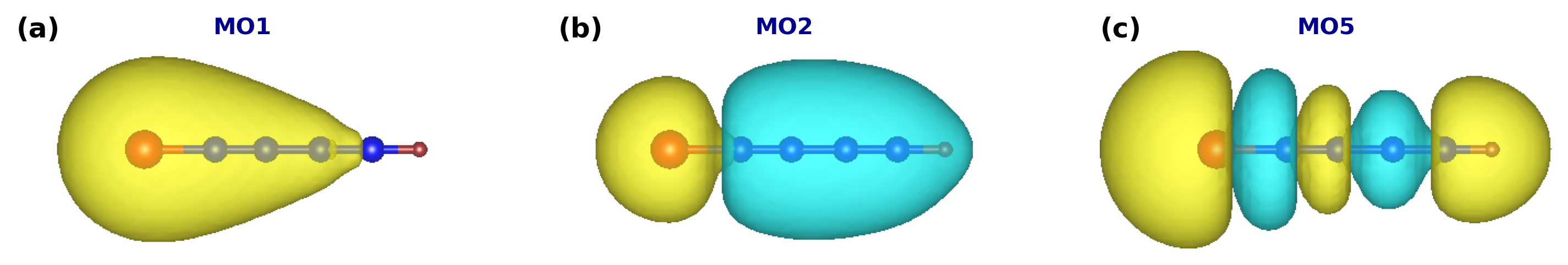}
\caption{Electronic density distribution of the inner valence orbitals of the selected molecule BrC\textsubscript{4}H highlighting the degree of localization on the Br site (the leftmost atom).}
\label{fig:si3}
\end{figure}

Fig.\ref{fig:si3} presents an overview of the degree of electron localization on the Br site in the selected MOs from which sudden ionization is selectively considered to initiate CM.  

\begin{figure}
\centering
\includegraphics[width=0.70\textwidth]{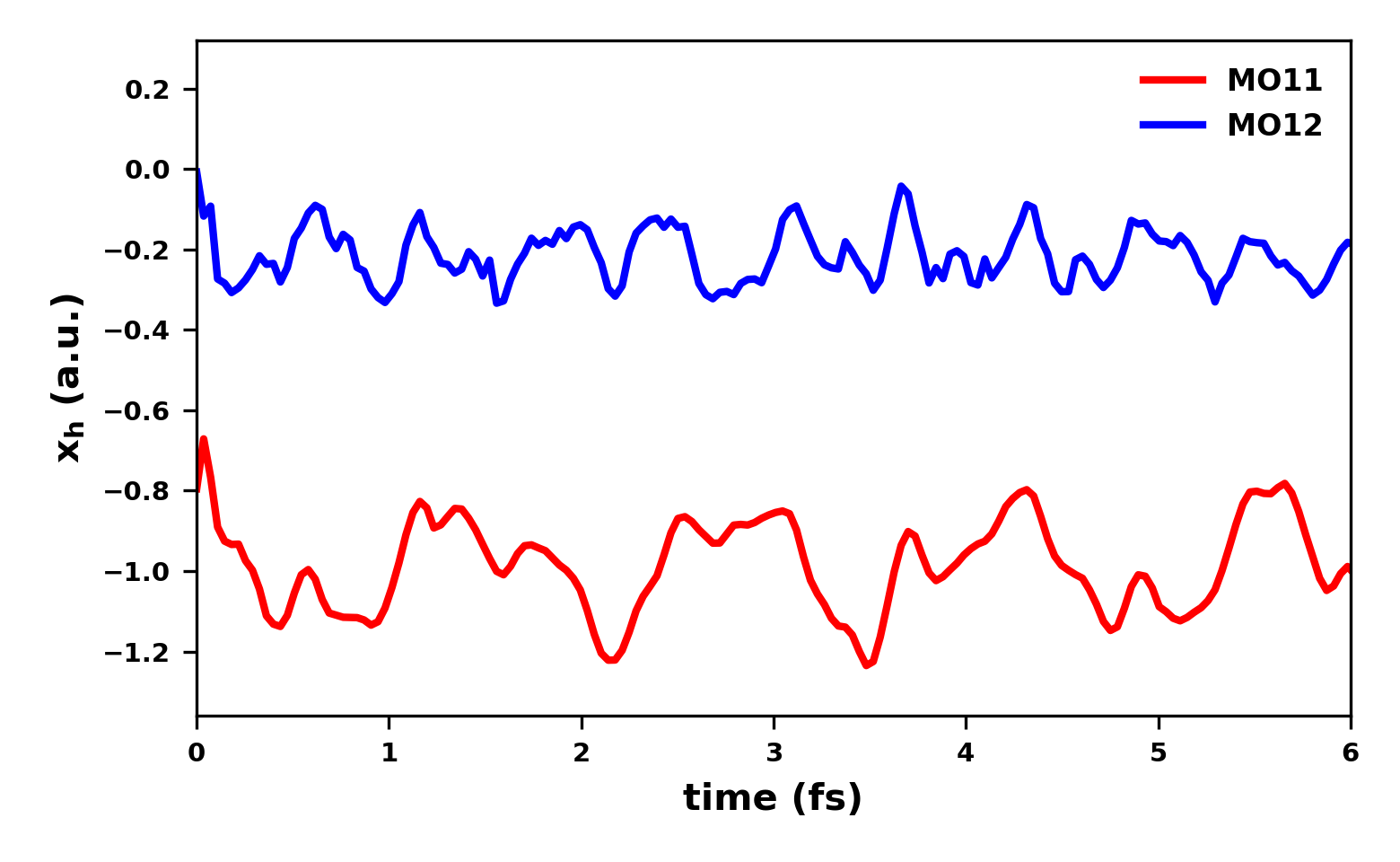}
\caption{Hole moments evolution as the CM advances from attosecond to few-femtosecond timescales when sudden ionization is initiated from the outer valence orbitals- HOMO (MO12) and HOMO-1 (MO11).}
\label{fig:si4}
\end{figure}

Fig.\ref{fig:si4} showcases hole moments timescales when sudden ionization is performed from the outer valence orbitals. Compared to the hole moments timescales presented in the main text, we notice a significant change in the CM period. For MO11, the CM period is $~\simeq1.3 fs$, whereas for MO12 it turns out to be $~\simeq0.6 fs$.

\subsubsection*{Semi-analytical reconstruction of hole moments}

\begin{table}\centering
\caption{Dominant coherent pairs from reconstructed hole moment}

\begin{tabular}{c cc cc}
\toprule
S. No. & Coherent Pairs $(I,J)$ & $D_{IJ}$  \\
\midrule
1 &  (12, 26)    & 0.07    \\
2 &  (11, 27)   & 0.07    \\
3 & (7, 14) & 0.03  \\
4 &  (8, 13)    & 0.03    \\
5 &  (12, 14)   & 2.69   \\
6 &  (11, 13)  & 2.69  \\
7 &  (9, 13)    & 0.01    \\
8 &   (10, 14)    & 0.01    \\
9 & (5, 15) & 1.41  \\
10 & (9, 27)  & 0.02 \\

\bottomrule
\end{tabular}
\label{table:1}
\end{table}

Table \ref{table:1} enlists 10 dominant coherent pairs $(I,J)$ in their decreasing order (in terms of their contribution; see the main text) for the reconstructed hole moment within the adiabatic local density approximation (aLDA). $D_{IJ}$ denotes the dipole matrix elements corresponding to each pair of states $(I,J)$. 







\end{document}